\documentclass[
  longbibliography,
  twocolumn,
  pre,
  amsmath,amssymb,
  aps
]{revtex4-2}
\usepackage[margin=2cm]{geometry}
\usepackage[T1]{fontenc}
\usepackage[utf8]{inputenc}
\usepackage{graphicx}
\usepackage{dcolumn}
\usepackage{bm}

\usepackage{amsmath}
\usepackage{mathptmx}
\allowdisplaybreaks[4]
\usepackage{amssymb}
\usepackage{xcolor}
\usepackage{tikz}
\usepackage{placeins}
\usepackage{etoolbox}
\usepackage{amsfonts} 
\usepackage[normalem]{ulem}
\usepackage{algorithm} 
\usepackage{algpseudocode}
\usepackage[colorlinks=true, allcolors=blue]{hyperref}

\begin{document}

\title{Competition between private and expressed opinions in binary choice: the $\alpha$-EPO $q$-voter model}

\author{Barbara Kami\'nska}
\affiliation{Department of Computational Social Science, Faculty of Management,
Wroc{\l}aw University of Science and Technology, Poland}

\author{Barbara Nowak}
\affiliation{Faculty of Pure and Applied Mathematics,
Wroc{\l}aw University of Science and Technology, Poland}

\author{Arkadiusz Lipiecki}
\affiliation{Department of Computational Social Science, Faculty of Management,
Wroc{\l}aw University of Science and Technology, Poland }

\author{Katarzyna Sznajd-Weron}
\email{katarzyna.weron@pwr.edu.pl}
\affiliation{Department of Computational Social Science, Faculty of Management,
Wroc{\l}aw University of Science and Technology, Poland}
\affiliation{Complexity Science Hub Vienna, Austria}

\date{\today}
\begin{abstract}
People often express opinions that differ from their privately held views, a phenomenon known in economy as preference falsification. Expressed-private opinion (EPO) models capture this by assigning each agent two dynamical variables: a private (internal) and an expressed (external) opinion. Within the nonlinear $q$-voter model, two EPO variants have been studied so far: with and without self-anticonformity. In both formulations, agents update private and expressed binary opinions,  one after another and at the same rate, which has led to two update schemes studied previously: AT (act then think), in which an agent first updates its expressed and then its private opinion, and TA (think then act), in which the order is reversed. To eliminate this ad hoc distinction and quantify the interplay between private and expressed opinions, we introduce the $\alpha$-EPO $q$-voter model with asynchronous updating -- in each elementary step, an agent updates its private opinion with probability $\alpha$ or its expressed opinion with complementary probability $1-\alpha$. We derive mean-field theory and, for the first time for EPO $q$-voter dynamics, a pair approximation, and validate them with Monte Carlo simulations on artificial and real organizational networks. 
Comparing the two model variants, we show that the collective outcome controlled by $\alpha$ strongly depends on self-anticonformity: with self-anticonformity the results are robust to $\alpha$, whereas without it $\alpha$ shifts the agreement-disagreement threshold and {can change the type of phase transition. In the mean-field limit this change occurs only for $q=3$, but the pair approximation reveals an additional low-connectivity regime in which both $\alpha$ and the average degree $k$ control the emergence and width of hysteresis also for larger influence groups.} 
\end{abstract}

\maketitle

\section{\label{sec:Introduction}Introduction}
It is well known that people do not always express their true preferences, a phenomenon known as preference falsification, because they may fear social disapproval, sanctions, or negative consequences arising from openly stating views that deviate from perceived social norms \cite{kuran_preference_1987}. 
{This phenomenon, often coupled with inaccurate convictions about the true beliefs of others, can bias polling and digital trace data~\cite{holbrook_2010, galesic_2021}. Beyond survey measurement, preference falsification and misperceived social norms can hinder grassroots environmental action~\cite{geiger_2016, mildenberger_2019, judge_2023} and even aggravate economic inequality~\cite{bursztyn_2020}. Hence, explicitly modeling private and expressed opinion layers is important not only from the perspective of fundamental understanding of opinion formation but also has practical potential, possibly allowing for more accurate polls, better public policy-making, and more effective management of collective action problems. The recent application of the voter model to predictive analysis of French elections~\cite{vendeville_2025} and the role of social norms in the observed public opinion shifts~\cite{bursztyn_egorov_2020} further highlight the need to study voter-like expressed-private opinion (EPO) models~\cite{dong_opinion_2024}; for a short and recent review, see~\cite{kaminska_impact_2025}.}

Here, we focus on a specific aspect of these models: the effect of unequal update rates for private and expressed opinions in a binary choice framework, which is  relevant to surprisingly complex problems \cite{schelling1973}. We study this effect within the nonlinear $q$-voter framework, one of the most widely used statistical-physics approaches to opinion dynamics \cite{Castellano2009,nyczka_phase_2012,Mobilia2015,Gradowski2020}. 
{This framework is particularly relevant here because it captures social-psychological mechanisms such as conformity, independence, and anticonformity, and provides a natural basis for modeling the interplay between private and expressed opinions; for a broader discussion of links between opinion-dynamics models and social psychology, see~\cite{jedrzejewski_statistical_2019}.}

There are various approaches to modeling the interplay between private and expressed (often called public) opinions. Some models assume that private opinions do not change over time \cite{masuda_can_2011, gaisbauer_dynamics_2020, manfredi_private-public_2020, llabres_partisan_2023}. By contrast, others allow both behavior and attitude to change at the same rate \cite{banisch_opinion_2019, ye_influence_2019, leon-medina_fakers_2020, jacob_polarization_2023, dong_opinion_2024, peng_adaptive_2025}. Finally, there are models in which the update rates for expressed and private opinions differ. An interesting example is the model proposed by \textcite{piras_social_2022}, in which the agents are generally more likely to change their behavior than their attitudes. However, when agents experience strong cognitive dissonance \cite{Festinger1957ADissonance}, whether they change their behavior or opinion depends on how often they have altered their opinions in the past. Another example, much closer to what we propose in this paper, is the concealed voter model \cite{gastner_consensus_2018}, in which agents change their expressed opinion by copying the opinion of a randomly selected neighbor (with probability $c$) or externalizing their internal opinion (with probability $e$), while the private opinion is updated by internalizing the expressed one (with probability $i$).

Despite the fact that $q$-voter models~\cite{Castellano2009,nyczka_phase_2012,Mobilia2015,Gradowski2020} typically follow an asynchronous updating scheme -- at each elementary time step a single voter updates its opinion -- the two $q$-voter-based EPO models \cite{jedrzejewski_think_2018, kaminska_impact_2025} assume that private and expressed opinions are updated in a quasi-synchronous manner. In this formulation, at each step a single voter updates both its private and public opinion in a specified order. Within this updating scheme, it has been shown that the stationary average private and expressed opinions (i.e., the collective-level aggregates) are independent of whether agents update their private opinion before their expressed opinion, or vice versa \cite{jedrzejewski_think_2018, kaminska_impact_2025}. It remains unclear, however, whether introducing rate asymmetry between private and expressed opinions alters the collective dynamics. We therefore extend these models by introducing an asynchronous update scheme, in which only one layer is updated at each step: private opinions are updated with probability $\alpha$, and expressed opinions with the complementary probability $1-\alpha$. We refer to these extensions as $\alpha$-EPO models. We study both variants considered in ~\cite{jedrzejewski_think_2018, kaminska_impact_2025} and analyze the resulting dynamics on synthetic and empirical networks using mean-field and pair approximation, as well as Monte Carlo simulations.

Studying both variants in parallel is particularly interesting because, as recently suggested, the older version of the model can be interpreted as incorporating a form of self-anticonformity \cite{jedrzejewski_think_2018}, whereas the newer version does not include this mechanism \cite{kaminska_impact_2025}. Furthermore, \textcite{jedrzejewski_think_2018} and \textcite{kaminska_impact_2025} considered their models only on a complete graph, corresponding to a fully connected society in which everyone knows everyone. For large social systems, this assumption is clearly unrealistic. Here, we move beyond the complete graph by studying $\alpha$-EPO $q$-voter models on networks using Monte Carlo (MC) simulations and pair approximation (PA) analysis.

\section{\label{sec:Model}The model}
We consider a system of $N$ agents, indexed by $i=1,\ldots,N$, each described by two binary variables: $S_i(t)\in\{-1,+1\}$, the expressed opinion of agent $i$ at time $t$ (visible to other agents), and $\sigma_i(t)\in\{-1,+1\}$, the private opinion of agent $i$ at time $t$ (not visible to others). The binary variables represent a choice between two alternatives, such as \textit{yes} or \textit{no}; \textit{for} or \textit{against}; \textit{adopt} or \textit{not adopt}, etc. Agents occupy the nodes of an undirected, unweighted network, whose topology specifies the interaction patterns within the system. For each agent $i$, let $\mathcal{N}(i)$ denote the set of its neighbors and let $k_i = |\mathcal{N}(i)|$ be the degree of node $i$; thus, agent $i$ can interact only with agents in $\mathcal{N}(i)$. In conformity events, the influence group (the $q$-panel) is formed by selecting uniformly at random, without replacement, $q$ neighbors of agent $i$, i.e.,a subset $\mathcal{Q} \subset \mathcal{N}(i)\;\wedge\;|\mathcal{Q}|=q$. In all analytical calculations and algorithmic description of the model we assume that every node in the network has at least $q$ neighbors.

Agents can update their opinions at both levels (expressed and private) due either to social influence from their neighbors or to the need to reduce internal dissonance, which arises when private and expressed opinions are in conflict. The updating rules for public and private opinions differ; however, at both levels we consider two mechanisms: independence and conformity. Independence at the expressed level means that an agent simply verbalizes its private opinion, i.e., its expressed opinion is set equal to the private one, thereby reducing the agent's internal dissonance. In contrast, independence at the private level corresponds to stochastic noise, as in many other versions of the nonlinear voter model~\cite{carro_noisy_2016, khalil_zealots_2018, khalil_noisy_2019, peralta_stochastic_2018, vieira_pair_2020}. Specifically, we use the formulation proposed in \cite{nyczka_phase_2012}: under independence, an agent flips its private opinion to the opposite one with probability $1/2$.

Conformity at both levels is driven by social pressure exerted by other agents. In our models, the source of this social influence is a group of $q$ randomly selected neighbors, called the $q$-panel. At the expressed level, conformity comprises two scenarios, depending on the agent’s initial state. If an agent is initially in harmony (i.e., its private and expressed opinions coincide), it is less susceptible to influence. Therefore, it adjusts its expressed opinion to match the group only if the $q$-panel is unanimous, that is, if all $q$ selected neighbors share the same expressed opinion. This type of behavior is called compliance~\cite{nail_proposal_2013}. By contrast, an agent who is in dissonance and thus experiences internal tension is more willing to change. In this case, it is enough that at least one member of the $q$-panel publicly expresses an opinion consistent with the agent’s private beliefs to encourage the agent to speak up, align its expressed opinion with its private one, and thereby relieve dissonance. This type of response to social influence is referred to as disinhibitory contagion~\cite{nail_proposal_2013}, a phenomenon in which breaking the norm by even a single individual incites others to act according to their true intentions. 

Conformity at the private level also requires a source of influence. Here, a key difference emerges between the models proposed by \textcite{jedrzejewski_think_2018} and \textcite{kaminska_impact_2025}, and this distinction carries over to the two variants considered in this paper. In the first case, following \textcite{jedrzejewski_think_2018}, an agent conforms to a unanimous $q$-panel regardless of its expressed opinion. Consequently, the agent may change its private opinion to match the group while still expressing the opposite view. We refer to this as self-anticonformity, since the agent, at the expressed level, disagrees not only with the group but also with its own private opinion. In the second variant, following \textcite{kaminska_impact_2025}, this scenario is prohibited. In other words, an agent conforms privately only if the $q$-panel is unanimous and its opinion matches the agent’s expressed one. This also implies that conformity on the private level cannot lead to dissonance.

\begin{figure}[h]
    \centering
    \includegraphics[width=\linewidth]{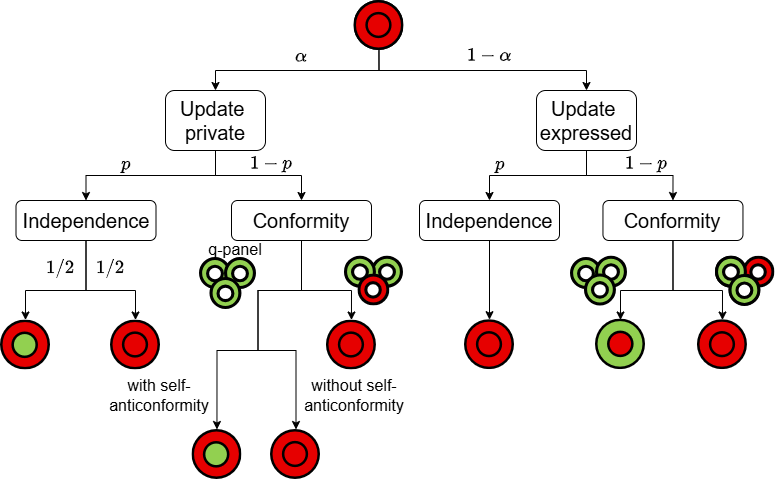}
    \caption{Visualization of a single update in the models with and without self-anticonformity (see the description directly above the bottom row in the left panel), for the case in which the focal agent is in harmony before the update.}
    \label{fig:model_harmony}
\end{figure}

\begin{figure}[h]
    \centering
    \includegraphics[width=\linewidth]{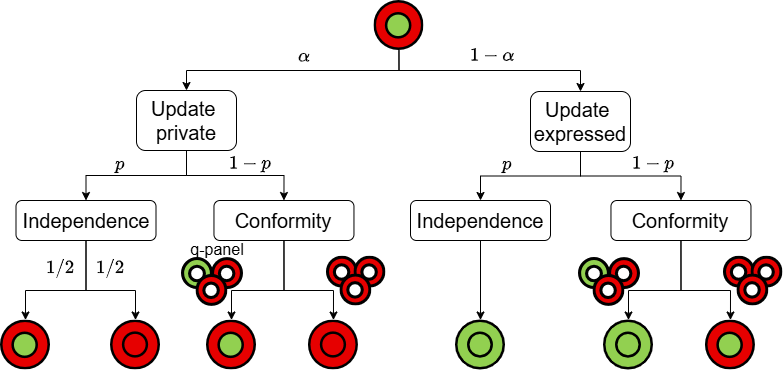}
    \caption{Visualization of a single update in the models with and without self-anticonformity (see the description directly above the bottom row in the left panel), for the case in which the focal agent is in dissonance before the update.}
    \label{fig:model_diss}
\end{figure}

In this paper, we introduce an extension that applies to both variants of the model described above, allowing for a more flexible relationship between updates at the expressed and private levels. The proposed extension captures the fact that a change in private opinion is not necessarily followed by a change in behavior, and vice versa: behavior may change without altering beliefs~\cite{kroesen_role_2018}. Both \textcite{jedrzejewski_think_2018, kaminska_impact_2025} assumed that during a single update, both expressed and private opinions are updated. They considered two update schemes: “act then think” (AT), in which the expressed opinion is updated first and the private one second, and “think then act” (TA), in which the private opinion is updated first and the expressed one second. Here, we assume that with probability $\alpha$ an agent updates its private opinion, and with complementary probability $1-\alpha$ it updates its expressed opinion. {We emphasize that an elementary update in the model should not be interpreted as a single real-life conversation, but rather as a coarse-grained opportunity for opinion revision. Moreover, selecting the private layer for an update, which occurs with probability $\alpha$, does not automatically imply a private-opinion change. The actual change occurs only if the corresponding conformity or independence condition is satisfied. Empirical studies provide quantitative evidence that social influence can induce opinion change at rates comparable to those considered in our model~\cite{chacoma_opinion_2015}.} The algorithm for an elementary update of the $\alpha$-EPO model, taking time $\Delta t$, is described by Algorithm \ref{alg:model}  and illustrated in Figs.~\ref{fig:model_harmony} and~\ref{fig:model_diss}.

Let us, following \cite{jedrzejewski_think_2018,kaminska_impact_2025}, use $\uparrow$ to denote the opinion $+1$ and $\downarrow$ to denote the opinion $-1$. To describe the state of a single agent at time $t$, we use the ordered pair $(S(t),\sigma(t))$, where the first element is the expressed opinion, followed by the private one. Accordingly, the set of possible agent states is
$X=(S,\sigma)\in \{\uparrow\uparrow,\uparrow\downarrow,\downarrow\uparrow,\downarrow\downarrow\}$.
The macroscopic state of the system at time $t$ can be described by the fractions of agents in each state, denoted by $c_X(t)$, defined as
\begin{flalign}
    c_X(t)=\frac{N_X(t)}{N}, \label{eq:def_concentration}
\end{flalign}
where $N_X(t)$ denotes the number of agents in state $X$ at time $t$, and $N$ is the total number of agents in the system. Since the total number of agents in the system is fixed, we have
\begin{flalign}
    c_{\uparrow \uparrow}(t) + c_{\uparrow \downarrow}(t) + c_{\downarrow \uparrow}(t) + c_{\downarrow \downarrow}(t) = 1. \label{eq:normalization}
\end{flalign}

To enable comparison with \textcite{jedrzejewski_think_2018} and \textcite{kaminska_impact_2025} instead of analyzing the concentrations of agents in all possible states separately, we consider the following quantities:
\begin{flalign}
c_S (t) = c_{\uparrow \uparrow} (t) + c_{\uparrow \downarrow} (t),
\label{eq:def_CS}
\end{flalign} being the fraction of agents with positive expressed opinion,  
\begin{flalign}
c_{\sigma}(t) = c_{\uparrow \uparrow}(t) + c_{\downarrow \uparrow}(t),
\label{eq:def_CSigma}
\end{flalign}
being the fraction of agents with positive private opinion, and
\begin{flalign}
d(t)= c_{\uparrow \downarrow}(t) + c_{\downarrow \uparrow}(t),
\label{eq:def_Cd}
\end{flalign}
being the fraction of agents in dissonance. 

\begin{algorithm}[H]
\caption{Elementary opinion update}
\label{alg:model}
\begin{algorithmic}[1]
\State Draw target agent $i \sim U\{1,\ldots,N\}$
\State Draw $r_{\alpha} \sim U(0,1)$

\If{$r_{\alpha} < \alpha$} \Comment{Update private opinion $\sigma_i$}
    \State Draw $r_p \sim U(0,1)$
    
    \If{$r_p < p$} \Comment{Independence}
        \State $\sigma_i(t+\Delta t) \gets (-1)^x \text{, where }x\sim \text{Bernoulli}\left(\frac{1}{2}\right)$
        
    \Else \Comment{Conformity}
        \State Draw random set $\mathcal{Q} \subset \mathcal{N}(i)$ such that $|\mathcal{Q}|=q$
        
        \If{\textbf{model with self-anticonformity}}
            \If{$\forall j \in \mathcal{Q},\; S_j(t)=-\sigma_i(t)$}
                \State $\sigma_i(t+\Delta t) \gets -\sigma_i(t)$
            \Else
                \State $\sigma_i(t+\Delta t) \gets \sigma_i(t)$
            \EndIf
            
        \ElsIf{\textbf{model without self-anticonformity}}
            \If{$\forall j \in \mathcal{Q},\; S_j(t)=S_i(t)$}
                \State $\sigma_i(t+\Delta t) \gets S_i(t)$
            \Else
                \State $\sigma_i(t+\Delta t) \gets \sigma_i(t)$
            \EndIf
        \EndIf
    \EndIf

\Else \Comment{Update expressed opinion $S_i$}
    \State Draw $r_p \sim U(0,1)$
    
    \If{$r_p < p$} \Comment{Independence}
        \State $S_i(t+\Delta t) \gets \sigma_i(t)$
        
    \Else \Comment{Conformity}
        \State  Draw random set $\mathcal{Q} \subset \mathcal{N}(i)$ such that $|\mathcal{Q}|=q$
        
        \If{$S_i(t) \neq \sigma_i(t)$} \Comment{Disinhibitory contagion}
            \If{$\exists j \in \mathcal{Q} \text{ such that } S_j(t)=\sigma_i(t)$}
                \State $S_i(t+\Delta t) \gets \sigma_i(t)$
            \Else
                \State $S_i(t+\Delta t) \gets S_i(t)$
            \EndIf
            
        \Else \Comment{Compliance}
            \If{$\forall j \in \mathcal{Q},\; S_j(t)=-S_i(t)$}
                \State $S_i(t+\Delta t) \gets -S_i(t)$
            \Else
                \State $S_i(t+\Delta t) \gets S_i(t)$
            \EndIf
        \EndIf
    \EndIf
\EndIf

\State $t \gets t+\Delta t$
\end{algorithmic}
\end{algorithm}

\section{\label{sec:Methods} Methods}
\subsection{Monte Carlo simulations}
We use Monte Carlo simulations both to validate our analytical calculations for complete and random graphs, as well as to extend the analysis to other topologies, including the empirical networks. To obtain smooth trajectories comparable with the mean-field and pair approximations (which effectively correspond to the thermodynamic limit), we simulate relatively large systems with $N=10^4$ and $N=10^5$ nodes. We also conduct MC simulations on empirical networks from organizations \cite{fire_organization_2016}. In the latter case, some nodes may have degree lower than $q$. In such cases we assume that the $q$-panel consists of all neighbors, i.e., for each agent $i$ we take $q_i = \min(k_i, q)$. 

In our MC simulations, one MC step (MCS) consists of $N$ elementary updates performed according to the algorithm described in Section~\ref{sec:Model} and illustrated in Figs.~\ref{fig:model_harmony} and \ref{fig:model_diss}. {Consequently, the Monte Carlo step should be regarded as a model time unit rather than a fixed physical unit of time; its empirical duration would have to be calibrated separately for a specific application.}We assume that agents are initially in harmony, i.e., $ \forall_i ; S_i(0) = \sigma_i(0)$. We use two types of initial conditions: ordered, $c_S(0) = c_\sigma(0) = 1$, and random, where $c_S(0) = c_\sigma(0) = 1/2$. After initialization, we equilibrate the system for 3000 MCS, and then perform time averaging over the subsequent 2000 MCS. The simulation codes are publicly available at \href{https://github.com/BarbaraKaminska/EPO-q-voter}{github.com/BarbaraKaminska/EPO-q-voter}
.

\subsection{Mean field approximation}
As stated above, one Monte Carlo step consists of $N$ elementary updates, hence $N\Delta t=1$, which implies $\Delta t = 1/N$. During a single elementary update, the fraction of agents in a given state can increase by $1/N$, decrease by $1/N$, or remain unchanged. We define the probabilities of these changes as
\begin{flalign}
    \gamma_X^+(t) &= P(c_X (t + 1/N) = c_X(t) + 1/N), \\
    \gamma_X^-(t) &= P(c_X (t + 1/N) = c_X(t) - 1/N), \\
    \gamma_X^0(t) &= P(c_X (t + 1/N) = c_X(t) ).
\label{eq:gamma}
\end{flalign}
This yields a system of four equations. Below, we write it in explicit form. To simplify notation, in what follows we omit the explicit time argument $t$ whenever no confusion may arise; in particular, we write $c_X$ instead of $c_X(t)$ (and analogously for $\gamma_X^\pm$, $S_i$, and $\sigma_i$). Time dependence will be stated explicitly only when needed.

\begin{widetext}
    \begin{flalign}
    \frac{dc_{\uparrow \uparrow}}{dt} &= c_{\uparrow\downarrow}\alpha\left[p/2+(1-p)c_S^q\right] + c_{\downarrow \uparrow}(1 - \alpha) \left[1 - (1-p)(1-c_S)^q\right] \nonumber \\ 
    &- c_{\uparrow\uparrow}\left\{1- \alpha\left[1-p/2 \, \boxed{-(1-p)(1-c_S)^q}\right] - (1 - \alpha) \left[1 - (1-p)(1-c_S)^q\right]\right\},
    \label{eq:dcdt_up_up} \\
    \frac{dc_{\uparrow \downarrow}}{dt} &= c_{\uparrow \uparrow}\alpha\left[p/2 \, \boxed{+(1-p)(1-c_S)^q}\right]  +c_{\downarrow \downarrow}(1-\alpha)(1-p)c_S^q     
    - c_{\uparrow\downarrow}\left\{1 - \alpha\left[1 - p/2 - (1-p)c_S^q\right] - (1-\alpha)(1-p)c_S^q \right\},
    \label{eq:dcdt_up_down} \\    
    \frac{dc_{\downarrow \uparrow}}{dt} &= c_{\uparrow\uparrow}(1-\alpha)(1-p)(1-c_S)^q 
    + c_{\downarrow\downarrow}\alpha\left[p/2 \, \boxed{+ (1-p)c_S^q}\right] \nonumber \\
    &- c_{\downarrow\uparrow}\left\{ 1- \alpha\left[1-p/2-(1-p)(1-c_S)^q\right] - (1-\alpha)(1-p)(1-c_S)^q  \right\},  
    \label{eq:dcdt_down_up}\\ 
    \frac{dc_{\downarrow \downarrow}}{dt} &= c_{\uparrow\downarrow}(1-\alpha)\left[1-(1-p)c_S^q\right] 
    + c_{\downarrow\uparrow}\alpha[p/2+(1-p)(1-c_S)^q] \nonumber \\ & - c_{\downarrow\downarrow} \left\{ 1 - \alpha\left[1 - p/2 \, \boxed{-(1-p)c_S^q}\right] - (1-\alpha)\left[1-(1-p)c_S^q\right] \right\}.
    \label{eq:dcdt_down_down}
\end{flalign}
\end{widetext}

In Eqs.\eqref{eq:dcdt_up_up}-\eqref{eq:dcdt_down_down}, the terms enclosed in frames indicate contributions that are absent in the variant without self-anticonformity. Consequently, the equations for the model without self-anticonformity are obtained by simply dropping the framed terms. These contributions correspond to updates in which private-level conformity leads to dissonance.

The stationary state, that we are interested in, is defined by 
\begin{flalign}
\frac{dc_X}{dt}=0
\end{flalign} for all $X$. Although the resulting system can be written in a fairly concise form, it must be solved numerically.

\subsubsection{Model with self-anticonformity}
Below we simplify the mean-field equations for the stationary states for the model with self-anticonformity. 

\begin{widetext}
    \begin{flalign}
    c_{\uparrow \uparrow} &= c_{\sigma}(1 - \alpha) \left[1 - (1-p)(1-c_S)^q\right]
    + c_{\uparrow \uparrow}\alpha\left[1-p/2 \, -(1-p)(1-c_S)^q\right] 
    + c_{\uparrow\downarrow}\alpha\left[p/2+(1-p)c_S^q\right],
    \label{eq:c_up_up_2018} \\
    c_{\uparrow \downarrow} &= (1-c_{\sigma})(1-\alpha)(1-p)c_S^q 
    + c_{\uparrow \uparrow}\alpha\left[p/2 + (1-p)(1-c_S)^q\right] 
    + c_{\uparrow\downarrow}\alpha\left[1 - p/2 - (1-p)c_S^q\right],
    \label{eq:c_up_down_2018} \\    
    c_{\downarrow \uparrow} &= c_{\sigma}(1-\alpha)(1-p)(1-c_S)^q 
    + c_{\downarrow\uparrow}\alpha\left[1-p/2-(1-p)(1-c_S)^q\right] 
    + c_{\downarrow\downarrow}\alpha\left[p/2 \, + (1-p)c_S^q\right],  
    \label{eq:c_down_up_2018}\\ 
    c_{\downarrow \downarrow} &= (1-c_{\sigma})(1-\alpha)\left[1-(1-p)c_S^q\right] 
    + c_{\downarrow\uparrow}\alpha[p/2+(1-p)(1-c_S)^q] 
    + c_{\downarrow\downarrow}\alpha\left[1 - p/2 \, -(1-p)c_S^q\right].
    \label{eq:c_down_down_2018}
\end{flalign}

\end{widetext}

We can proceed analogously to \textcite{jedrzejewski_think_2018}, that is we add \eqref{eq:c_up_up_2018} and \eqref{eq:c_up_down_2018}, then \eqref{eq:c_up_up_2018} and \eqref{eq:c_down_up_2018}. Then, using Eqs. \eqref{eq:def_CS} and \eqref{eq:def_CSigma} we finally obtain 
\begin{flalign}
    c_S (1-\alpha) &= c_{\sigma}(1-\alpha)\left[1-(1-p)\left[(1-c_S)^q+c_S^q\right]\right] \nonumber \\
    &+(1-\alpha) (1-p)c_S^q, \label{eq:system_2018_cs}\\
    c_{\sigma} \alpha &= c_{\sigma}\alpha\left[1 - p/2 - (1-p)(1-c_S)^q\right] \nonumber \\
    &+ (1-c_\sigma)\alpha\left[p/2 + (1-p) c_S^q\right]. \label{eq:system_2018_csigma}
\end{flalign}

Hence for $\alpha \neq 0$ and $\alpha \neq 1$ we obtain exactly the same system of equations as \textcite{jedrzejewski_think_2018}, namely 
\begin{flalign}
    c_S & = c_{\sigma}\left[1-(1-p)\left[(1-c_S)^q+c_S^q\right]\right]+(1-p)c_S^q,     \label{eq:final_system_2018}   \\
    c_{\sigma} & = \frac{p/2+ (1-p)c_S^q}{p+(1-p)\left[(1-c_S)^q+c_S^q\right]}. \label{eq:final_system_2018a} 
\end{flalign}
This means that, for the model with self-anticonformity, the parameter $\alpha$ does not affect the stationary values of the average private and expressed opinions as long as $\alpha \in (0,1)$.

Let us notice that for $\alpha = 0$, Eq.~\eqref{eq:system_2018_csigma} becomes an identity, and Eq.~\eqref{eq:system_2018_cs} is the same as \eqref{eq:final_system_2018}. Recalling that for $\alpha = 0 $, agents update only their expressed opinion, the aforementioned observations mean that the fraction of agents with a positive expressed opinion depends on the initial fraction of agents with a positive private opinion. This dependence can be derived by solving eq. \eqref{eq:final_system_2018} for $c_S$. Analogously, Eq.~\eqref{eq:system_2018_cs} becomes an identity equation for $\alpha = 1$. In this case, agents update only their private opinions, and we can solve Eq.~\eqref{eq:final_system_2018a} for $c_\sigma$ by plugging in the independence probability $p$ and the initial fraction of public opinion $c_S$.

\subsubsection{Model without self-anticonformity}
As stated above, for the model without self-anticonformity, the system is reduced analogously, we simply remove the framed terms, which yields:

\begin{widetext}
\begin{flalign}
    c_{\uparrow \uparrow} &=c_{\sigma} \left[1 - (1-p)(1-c_S)^q\right](1 - \alpha) + c_{\uparrow \uparrow}\alpha\left[1-p/2\right] + c_{\uparrow\downarrow}\alpha\left[p/2+(1-p)c_S^q\right], \label{eq:c_up_up_2025} \\
    c_{\uparrow \downarrow} &= (1-c_{\sigma})(1-\alpha)(1-p)c_S^q + c_{\uparrow \uparrow}\alpha p/2 + c_{\uparrow\downarrow}\alpha\left[1 - p/2 - (1-p)c_S^q\right], \label{eq:c_up_down_2025} \\    
    c_{\downarrow \uparrow} &= c_{\sigma}(1-\alpha)(1-p)(1-c_S)^q + c_{\downarrow\uparrow}\alpha\left[1-p/2-(1-p)(1-c_S)^q\right] + c_{\downarrow\downarrow}\alpha p/2,  \label{eq:c_down_up_2025}\\ 
    c_{\downarrow \downarrow} &= (1-c_{\sigma})(1-\alpha)\left[1-(1-p)c_S^q\right] + c_{\downarrow\uparrow}\alpha\left[p/2+(1-p)(1-c_S)^q\right] + c_{\downarrow\downarrow}\alpha\left[1 - p/2 \right].  \label{eq:c_down_down_2025}
\end{flalign}

\end{widetext}

By adding \eqref{eq:c_up_up_2025} and \eqref{eq:c_up_down_2025} we obtain 
\begin{flalign}
    c_S (1-\alpha) &= c_{\sigma}(1-\alpha)\left[1-(1-p)\left[(1-c_S)^q+c_S^q\right]\right] \nonumber \\ 
    & +(1-\alpha) (1-p)c_S^q.
    \label{eq:cs_B}
\end{flalign}
It is worth noting that for $\alpha \ne 1$  this simplifies to Eq. \eqref{eq:final_system_2018}. However, adding Eqs.~\eqref{eq:c_up_up_2025} and \eqref{eq:c_down_up_2025} does not lead to such a simple expression: the result cannot be written solely as a function of $c_S$, $c_\sigma$, and the model parameters $q$, $p$, and $\alpha$.

Again, let us briefly consider the extreme cases of $\alpha$ equal to 0 and 1. For $\alpha = 0$, the result is the same as for the model with self-anticonformity, as in this scenario agents update only their expressed opinion, which evolves according to the same algorithm in both variants of the model. When $\alpha = 1$, Eq.~\eqref{eq:cs_B} becomes an identity, and consequently we cannot derive simple dependence of $c_\sigma$ on $c_S$. For $\alpha = 1$, Eqs.~\eqref{eq:c_up_up_2025}-\eqref{eq:c_down_down_2025} include distinct terms containing $c_{\uparrow\downarrow}$ and $c_{\downarrow\uparrow}$, which cannot be reduced to depend solely on $c_S$ and $c_\sigma$.

\subsection{Pair approximation}

Pair approximation is an analytical approach that relaxes the assumption of a completely homogeneous agent system by accounting for dynamical correlations between pairs of agents in different states. It has been successfully applied to many versions of the $q$-voter model, in which agents are described by a single opinion \cite{Jedrzejewski2017PairNetworks,Gradowski2020,vieira_pair_2020,jedrzejewski_pair_2022,Ramirez2024}, but not to the EPO $q$-voter models. Within this framework, the evolution equations for the state concentrations in the model with self-anticonformity are given by the following system (for the model without self-anticonformity, the framed terms are omitted): 
\begin{widetext}
\begin{flalign}
    \frac{dc_{\uparrow\uparrow}}{dt} & = (1-\alpha)\left[pc_{\downarrow\uparrow} + (1-p)\left\{c_{\downarrow\uparrow}\left(1 - \left( \theta_{\downarrow\uparrow}^{\downarrow\uparrow} + \theta_{\downarrow\uparrow}^{\downarrow\downarrow}\right)^q \right) - c_{\uparrow\uparrow}\left(\theta_{\uparrow\uparrow}^{\downarrow\uparrow} + \theta_{\uparrow\uparrow}^{\downarrow\downarrow}\right)^q
    \right\} \right] \nonumber \\ 
& +\alpha\left[\frac{p}{2}(c_{\uparrow\downarrow} - c_{\uparrow\uparrow}) + (1-p)\left\{c_{\uparrow\downarrow}\left(\theta_{\uparrow\downarrow}^{\uparrow\uparrow} + \theta_{\uparrow\downarrow}^{\uparrow\downarrow}\right)^q \boxed{ - c_{\uparrow\uparrow}\left(\theta_{\uparrow\uparrow}^{\downarrow\uparrow} + \theta_{\uparrow\uparrow}^{\downarrow\downarrow}\right)^q} \right\}\right] \nonumber \\
\frac{dc_{\uparrow\downarrow}}{dt} & = (1-\alpha)\left[-p c_{\uparrow\downarrow} + (1-p)\left\{c_{\downarrow\downarrow}\left(\theta_{\downarrow\downarrow}^{\uparrow\uparrow} + \theta_{\downarrow\downarrow}^{\uparrow\downarrow}\right)^q - c_{\uparrow\downarrow}\left(1 - \left(\theta_{\uparrow\downarrow}^{\uparrow\uparrow} + \theta_{\uparrow\downarrow}^{\uparrow\downarrow}\right)^q\right)\right\}\right] \nonumber \\
& + \alpha\left[\frac{p}{2}(c_{\uparrow\uparrow} - c_{\uparrow\downarrow}) + (1-p)\left\{ \boxed{ c_{\uparrow\uparrow}\left(\theta_{\uparrow\uparrow}^{\downarrow\uparrow} + \theta_{\uparrow\uparrow}^{\downarrow\downarrow}\right)^q } - c_{\uparrow\downarrow}\left(\theta_{\uparrow\downarrow}^{\uparrow\uparrow} + \theta_{\uparrow\downarrow}^{\uparrow\downarrow}\right)^q\right\}\right]\nonumber \\
    \frac{dc_{\downarrow\uparrow}}{dt} & = (1-\alpha)\left[-p c_{\downarrow\uparrow} + (1-p)\left\{ c_{\uparrow\uparrow}\left(\theta_{\uparrow\uparrow}^{\downarrow\uparrow} + \theta_{\uparrow\uparrow}^{\downarrow\downarrow}\right)^q - c_{\downarrow\uparrow}\left(1 - \left(\theta_{\downarrow\uparrow}^{\downarrow\uparrow} + \theta_{\downarrow\uparrow}^{\downarrow\downarrow}\right)^q\right) \right\} \right] \nonumber \\
    & + \alpha \left[\frac{p}{2}(c_{\downarrow\downarrow} - c_{\downarrow\uparrow}) + (1-p)\left\{ \boxed{ c_{\downarrow\downarrow}\left(\theta_{\downarrow\downarrow}^{\uparrow\uparrow} + \theta_{\downarrow\downarrow}^{\uparrow\downarrow}\right)^q  } - c_{\downarrow\uparrow}\left(\theta_{\downarrow\uparrow}^{\downarrow\uparrow} + \theta_{\downarrow\uparrow}^{\downarrow\downarrow}\right)^q \right\}\right]\nonumber\\
\frac{dc_{\downarrow\downarrow}}{dt} & = (1-\alpha)\left[p c_{\uparrow\downarrow} + (1-p)\left\{c_{\uparrow\downarrow}\left(1 - \left(\theta_{\uparrow\downarrow}^{\uparrow\uparrow} + \theta_{\uparrow\downarrow}^{\uparrow\downarrow}\right)^q\right) - c_{\downarrow\downarrow}\left(\theta_{\downarrow\downarrow}^{\uparrow\uparrow} + \theta_{\downarrow\downarrow}^{\uparrow\downarrow}\right)^q\right\}\right] \nonumber \\
& +\alpha\left[\frac{p}{2}(c_{\downarrow\uparrow} - c_{\downarrow\downarrow}) + (1-p)\left\{c_{\downarrow\uparrow}\left(\theta_{\downarrow\uparrow}^{\downarrow\uparrow} + \theta_{\downarrow\uparrow}^{\downarrow\downarrow}\right)^q \boxed{ - c_{\downarrow\downarrow}\left(\theta_{\downarrow\downarrow}^{\uparrow\uparrow} + \theta_{\downarrow\downarrow}^{\uparrow\downarrow}\right)^q} \right\}\right] \label{eq:PA_4_states}
\end{flalign}
and thus, according to definitions \eqref{eq:def_CS} and \eqref{eq:def_CSigma},  the evolution of aggregate concentrations of positive public and positive private opinions can be expressed as:
\begin{flalign}
    \frac{dc_S}{dt} & = (1-\alpha)(1-p)\left[ \left\{c_{\downarrow\uparrow}\left(1 - \left( \theta_{\downarrow\uparrow}^{\downarrow\uparrow} + \theta_{\downarrow\uparrow}^{\downarrow\downarrow}\right)^q \right) + c_{\downarrow\downarrow}\left(\theta_{\downarrow\downarrow}^{\uparrow\uparrow} + \theta_{\downarrow\downarrow}^{\uparrow\downarrow}\right)^q   - c_{\uparrow\downarrow}\left(1 - \left( \theta_{\uparrow\downarrow}^{\uparrow\uparrow} + \theta_{\uparrow\downarrow}^{\uparrow\downarrow}\right)^q \right) - c_{\uparrow\uparrow}\left(\theta_{\uparrow\uparrow}^{\downarrow\uparrow} + \theta_{\uparrow\uparrow}^{\downarrow\downarrow}\right)^q\right\}\right] \nonumber \\
    & + (1-\alpha) p \left(c_{\downarrow\uparrow} - c_{\uparrow\downarrow}\right), \nonumber \\ 
    \frac{dc_\sigma}{dt} & = \alpha (1-p)\left\{\boxed{c_{\downarrow\downarrow}\left(\theta_{\downarrow\downarrow}^{\uparrow\uparrow} + \theta_{\downarrow\downarrow}^{\uparrow\downarrow}\right)^q } + c_{\uparrow\downarrow}\left(\theta_{\uparrow\downarrow}^{\uparrow\uparrow} + \theta_{\uparrow\downarrow}^{\uparrow\downarrow}\right)^q - c_{\downarrow\uparrow}\left(\theta_{\downarrow\uparrow}^{\downarrow\uparrow} + \theta_{\downarrow\uparrow}^{\downarrow\downarrow}\right)^q  \boxed{- c_{\uparrow\uparrow}\left(\theta_{\uparrow\uparrow}^{\downarrow\uparrow} + \theta_{\uparrow\uparrow}^{\downarrow\downarrow}\right)^q} \right\} \nonumber \\ & + \alpha \frac{p}{2}\left(c_{\downarrow\downarrow} + c_{\uparrow\downarrow} - c_{\downarrow\uparrow} - c_{\uparrow\uparrow} \right) \label{eq:PA_c_S_c_sigma}
\end{flalign}
\end{widetext}
where $\theta_X^Y$ denotes the probability that a randomly selected neighbor of a target agent in state $X$ is in state $Y$. In pair approximation, this probability is modeled via concentration of edges between all possible agent states:
\begin{flalign}
 \theta_X^Y = \frac{e_{X}^{Y}}{e_{X}^{\uparrow\uparrow} + e_{X}^{\uparrow\downarrow} + e_{X}^{\downarrow\uparrow}+ e_{X}^{\downarrow\downarrow}},
\end{flalign}
with $e_X^Y := E_X^Y /E$, where $E_X^Y$ is the number of directed edges starting at an agent in state $X$ and ending at an agent is state $Y$; and $E$ is the total number of edges in the system. Since we consider undirected networks, $e_{X}^{Y} = e_{Y}^{X}$ holds for any $X$ and $Y$, which requires us to track only 10 unique edge concentrations. Additionally, exploiting the fact that the sum of concentrations must equate to 1, a total of 9 independent edge concentrations have to be considered. The time evolution of edge concentrations are in general described by the following equation 
\begin{flalign}
     \frac{d e_{X}^{Y}}{dt} = \frac{1}{\langle k \rangle} \sum_{Z \neq W} c_Z \sum_{\kappa} P(\kappa \mid Z) f^{Z \rightarrow W}(\kappa) \Delta E_{X}^{Y}\mid^{Z \rightarrow W}(\kappa), 
     \label{eq:edge_evolution}
\end{flalign}
where $\langle k \rangle$ is the global average node degree, ${\kappa = [k_{\uparrow\uparrow},k_{\uparrow\downarrow}, k_{\downarrow\uparrow}, k_{\downarrow\downarrow}]}$ is a vector of state populations in the direct neighborhood, denoting the number of neighbors in each state; $P(\kappa\mid Z)$ is the probability that an agent in state $Z$ has neighborhood $\kappa$; $f^{Z \rightarrow W}(\kappa)$ is the probability of an agent changing from state $Z$ to $W$ conditional on its neighborhood $\kappa$; and $\Delta E_{X}^{Y}\mid^{Z \rightarrow W}(\kappa)$ is the elementary change in the number of edges from agents in state $X$ to agents in $Y$, which occurs when an agent changes from $Z$ to $W$. The latter is given as

\begin{flalign}
     \Delta E_{X}^{Y}\mid^{Z \rightarrow W}(\kappa) = k_Y(\delta_{W, X} - \delta_{Z, X}) + k_X(\delta_{W, Y} - \delta_{Z, Y}),
     \label{eq:edge:delta}
\end{flalign}
where $\delta_{i, j}$ is the Kronecker delta. 

Let us explicitly list the flipping probabilities for all possible transitions between agent states. Most of these probabilities are identical in the models with and without self-anticonformity and will be presented later. The only differences occur for the transitions $\uparrow\uparrow \rightarrow \uparrow\downarrow$ and $\downarrow\downarrow \rightarrow \downarrow\uparrow$, so we start with these cases.

For the model with self-anticonformity:
\begin{flalign}
         f^{\uparrow\uparrow \rightarrow \uparrow\downarrow} \left(\kappa \right) &= \alpha \left(1 - p\right) \binom{k_{\downarrow\uparrow}+k_{\downarrow\downarrow}}{q}\frac{1}{\binom{ k }{q}}\mathbf{1}_{k_{\downarrow\uparrow} + k_{\downarrow\downarrow} \geqslant q} +  \frac{1}{2}\alpha p\nonumber \\
         &= \alpha \left(1 - p\right)\frac{\left(k_{\downarrow\uparrow}+k_{\downarrow\downarrow}\right)!\left(k-q\right)!}{\left(k_{\downarrow\uparrow}+k_{\downarrow\downarrow} - q\right)!k!}\mathbf{1}_{k_{\downarrow\uparrow} + k_{\downarrow\downarrow} \geqslant q} \nonumber \\ 
         &+ \frac{1}{2}\alpha p ,\label{flipping_prob1a}\\
         f^{\downarrow\downarrow \rightarrow \downarrow\uparrow} \left(\kappa \right) &= \alpha \left(1 - p\right) \frac{\left(k_{\uparrow\uparrow}+k_{\uparrow\downarrow}\right)!\left(k-q\right)!}{\left(k_{\uparrow\uparrow}+k_{\uparrow\downarrow} - q\right)!k!}\mathbf{1}_{k_{\uparrow\uparrow} + k_{\uparrow\downarrow} \geqslant q} \nonumber \\ & +\frac{1}{2}\alpha p \label{flipping_prob2a}, 
\end{flalign}

while without self-anticonformity we get:
\begin{flalign}
         f^{\uparrow\uparrow \rightarrow \uparrow\downarrow} \left(\kappa \right) &= \frac{1}{2}\alpha p, \label{flipping_prob1b}\\
         f^{\downarrow\downarrow \rightarrow \downarrow\uparrow} \left(\kappa \right) &= \frac{1}{2}\alpha p \label{flipping_prob2b}.
\end{flalign}

The flipping probabilities of the remaining transitions are the same for both models:
\begin{flalign}
f^{\uparrow\uparrow \rightarrow \downarrow\uparrow} \left(\kappa \right) &= \left(1 - \alpha\right)\left(1 - p\right) \frac{\left(k_{\downarrow\uparrow}+k_{\downarrow\downarrow}\right)!\left(k-q\right)!}{\left(k_{\downarrow\uparrow}+k_{\downarrow\downarrow} - q\right)!k!}\mathbf{1}_{k_{\downarrow\uparrow} + k_{\downarrow\downarrow} \geqslant q}, \label{flipping_prob3}\\
     f^{\uparrow\downarrow \rightarrow \uparrow\uparrow} \left(\kappa \right) 
     &= \alpha \left(1 - p\right)\frac{\left(k_{\uparrow\uparrow}+k_{\uparrow\downarrow}\right)!\left(k-q\right)!}{\left(k_{\uparrow\uparrow}+k_{\uparrow\downarrow} - q\right)!k!}\mathbf{1}_{k_{\uparrow\uparrow} + k_{\uparrow\downarrow} \geqslant q} \nonumber \\ & + \frac{1}{2}\alpha p, \label{flipping_prob4} \\
 f^{\uparrow\downarrow \rightarrow \downarrow\downarrow} \left(\kappa \right) &=\left(1 - \alpha\right) \nonumber \\ & \times \left[1 - \left(1 - p\right) \frac{\left(k_{\uparrow\uparrow}+k_{\uparrow\downarrow}\right)!\left(k-q\right)!}{\left(k_{\uparrow\uparrow}+k_{\uparrow\downarrow} - q\right)!k!}\mathbf{1}_{k_{\uparrow\uparrow} + k_{\uparrow\downarrow} \geqslant q}\right], \label{flipping_prob5}\\
    f^{\downarrow\uparrow \rightarrow \downarrow\downarrow} \left(\kappa \right)&= \alpha \left(1 - p\right) \frac{\left(k_{\downarrow\uparrow}+k_{\downarrow\downarrow}\right)!\left(k-q\right)!}{\left(k_{\downarrow\uparrow}+k_{\downarrow\downarrow} - q\right)!k!}\mathbf{1}_{k_{\downarrow\uparrow} + k_{\downarrow\downarrow} \geqslant q} \nonumber \\ & + \frac{1}{2} \alpha p,\label{flipping_prob6}\\
 f^{\downarrow\uparrow \rightarrow \uparrow\uparrow} \left(\kappa \right) &=\left(1 - \alpha\right)\nonumber \\ & \times \left[1 - \left(1 - p\right) \frac{\left(k_{\downarrow\uparrow}+k_{\downarrow\downarrow}\right)!\left(k-q\right)!}{\left(k_{\downarrow\uparrow}+k_{\downarrow\downarrow} - q\right)!k!}\mathbf{1}_{k_{\downarrow\uparrow} + k_{\downarrow\downarrow} \geqslant q}\right],\label{flipping_prob7}\\
     f^{\downarrow\downarrow \rightarrow \uparrow\downarrow} \left(\kappa \right) &= \left(1 - \alpha\right)\left(1 - p\right) \frac{\left(k_{\uparrow\uparrow}+k_{\uparrow\downarrow}\right)!\left(k-q\right)!}{\left(k_{\uparrow\uparrow}+k_{\uparrow\downarrow} - q\right)!k!}\mathbf{1}_{k_{\uparrow\uparrow} + k_{\uparrow\downarrow} \geqslant q}.\label{flipping_prob8}
 \end{flalign}

Finally, substituting Eqs.~\eqref{eq:edge:delta}-\eqref{flipping_prob8} into Eq.~\eqref{eq:edge_evolution} allows us to obtain a system of differential equations that does not depend on the node degree distribution, but only on the average degree $\langle k \rangle_X$. The average node degree can be inferred from edge and state concentrations according to:
\begin{equation}
    \langle k \rangle_X = \frac{\sum_{Y}e_{X}^{Y}}{c_X}\langle k \rangle.
    \label{eq:avg_degree_annealed}
\end{equation}

As in other homogeneous pair-approximation treatments of 
$q$-voter models \cite{jedrzejewski_pair_2022, lipiecki_when_2025}, correlations between node degree and state do not emerge unless they are present in the initial condition. Hence, we can set $\langle k \rangle_X = \langle k \rangle$ for all $X$ and extract the opinion concentration from Eq.~\eqref{eq:avg_degree_annealed} according to:
\begin{equation}
    c_X = \sum_{Y}e_{X}^{Y},
\end{equation}
allowing us to fully describe the evolution of the system with $9$ edge concentrations. Detailed formulas for the differential equations describing the time evolution of the concentrations of all edges are included in the source codes, publicly available at \href{https://github.com/BarbaraKaminska/EPO-q-voter}{github.com/BarbaraKaminska/EPO-q-voter}.

\section{\label{sec:Results}Results}

\subsection{\label{Results_MFA}Mean-field results} 
We begin by discussing the results for a complete graph, where all agents are mutually connected and can interact with each other. All-to-all interactions are consistent with the homogeneity assumed in the mean-field approximation, so in this setting the analytical predictions are expected to coincide with Monte Carlo simulations. We focus on the stationary values of three macroscopic quantities: the fraction of agents with positive expressed opinion, $c_S$, the fraction of agents with positive private opinion, $c_{\sigma}$, and the fraction of agents in dissonance, $d$ (see Eqs.~\eqref{eq:def_CS}-\eqref{eq:def_Cd}). We analyze these quantities as functions of the independence probability $p$ for the two versions of the $\alpha$-EPO model: with  and without self-anticonformity.

As shown in Fig.~\ref{fig:alpha_dependence}, both models exhibit a phase transition between an ordered phase (agreement), in which the majority of agents share the same opinion ($c_S \neq 0.5$, $c_\sigma \neq 0.5$), and a disordered phase (disagreement), in which opinions are evenly split ($c_S = 0.5$, $c_\sigma = 0.5$). The location of the critical point $p^*$ and the type of the transition differ between the two $\alpha$-EPO variants. Nevertheless, within each variant, $c_S(p)$ and $c_\sigma(p)$ undergo the transition at the same critical point, and the transition is of the same type for both quantities. In the ordered phase ($p<p^*$), the majority opinion is more pronounced at the expressed level than at the private level. This result is consistent with pluralistic ignorance, i.e., a mismatch in which privately held views are more diverse than publicly expressed ones because individuals misperceive the prevailing norm and adjust their public behavior accordingly \cite{Miller2023}.

The role of $\alpha$ differs qualitatively between the two variants of the $\alpha$-EPO model. In the version with self-anticonformity, the stationary phase diagrams $c_S(p)$ and $c_\sigma(p)$ are independent of $\alpha$ and therefore coincide with those reported in \textcite{jedrzejewski_think_2018} (top row in Fig.~\ref{fig:alpha_dependence}). This is consistent with Eqs.~\eqref{eq:final_system_2018}--\eqref{eq:final_system_2018a}, which, for $0<\alpha<1$, do not contain $\alpha$. In this variant, $\alpha$ affects only the stationary dissonance $d$, which increases with $\alpha$. This agrees with \textcite{jedrzejewski_think_2018}, where dissonance was also the only quantity that distinguished the two update schemes: act then think (AT), in which the expressed opinion was updated first and the private opinion second, and think then act (TA), in which the private opinion was updated first and the expressed opinion second. 

This robustness with respect to $\alpha$ is lost in the model without self-anticonformity (bottom row in Fig.~\ref{fig:alpha_dependence}). In this variant, the stationary curves $c_S(p)$, $c_\sigma(p)$, and $d(p)$ vary with $\alpha$. An intuitive explanation follows from the fact that private conformity is conditional on the expressed state in this model: even if the $q$-panel is unanimous, a private update is allowed only when the unanimous opinion of the panel coincides with the target’s expressed opinion $S_i$. Hence, the expressed layer effectively acts as a filter that can block private conformity events. As a consequence, changing $\alpha$ affects not only the relative frequency of updates at the private and expressed levels, but also the effectiveness of private conformity. 

\begin{figure}[htb]
    \centering
    \includegraphics[width=\linewidth]{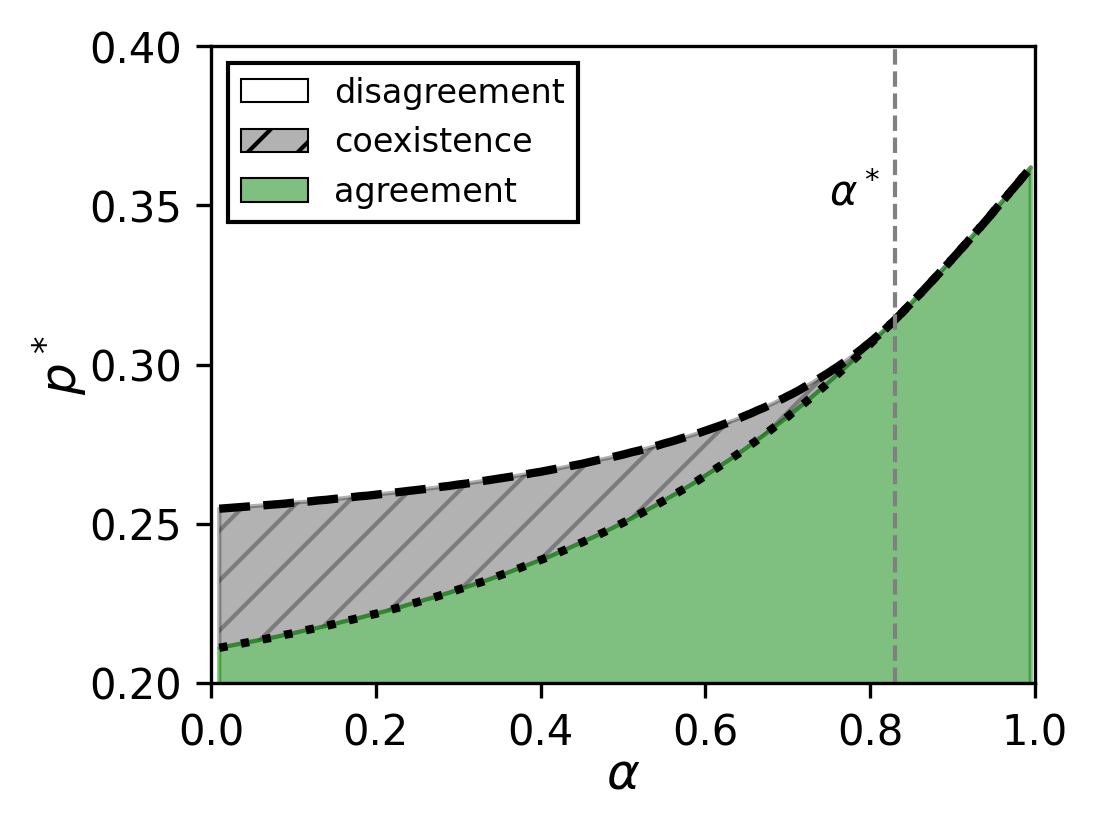}
    \caption{{\textbf{Phase diagram for $q=3$.} Dotted lines correspond to the lower spinodal, while dashed lines correspond to the upper spinodal. The two spinodals overlap for $\alpha \geq \alpha^*$, indicating a continuous phase transition. The agreement (ordered) phase is marked by a green solid fill, the coexistence region by a gray striped fill, and the disagreement (disordered) phase by a white no-fill region.}}
    \label{fig:phase_diagram}
\end{figure}

Varying $\alpha$ can modify not only the location of the critical point $p^*$ but also the type of phase transition. {Within the mean-field approximation, which corresponds to the complete graph structure, this qualitative effect is restricted to the borderline case $q=3$. For $q=3$, there exists a critical value $0.8 \leq \alpha^* \leq 0.9$ above which the discontinuous transition becomes continuous, as shown in Fig.~\ref{fig:phase_diagram}. For $q \leq 2$ the transition remains continuous, whereas for $q \geq 4$ it remains discontinuous for all considered values of $\alpha$ (see Fig.~\ref{fig:alpha_dependence_q_2_4} in Appendix~\ref{sec:appendix}). However, as shown below, this mean-field picture is modified on sparse networks, where the average degree $k$ becomes an additional control parameter.}

\subsection{\label{Results_PA}Pair approximation results}
We now extend the analysis beyond complete graphs and account for network topology using the pair approximation. Within this approach we solve both $\alpha$-EPO variants, with and without strategic self-anticonformity, and compare the resulting phase behavior with the mean-field predictions. {Overall, the pair approximation confirms the qualitative distinction between the two model variants found in the mean-field analysis, but it also reveals an important effect that is absent in the complete-graph limit. In the model with self-anticonformity, the stationary values of global concentrations of private and public opinions are independent of $\alpha$ for $0<\alpha<1$, consistently with the mean-field result. By contrast, in the model without self-anticonformity, the interplay between $\alpha$ and the average degree $k$ becomes essential, especially for sparse networks.}

{For discontinuous transitions, the hysteresis region is delimited by two spinodals, $p^*_{low}$ and $p^*_{up}$, which mark the limits of stability of the ordered and disordered phases. Figure~\ref{fig:spinodals_PA} shows how these spinodals depend on $\alpha$ for different influence-group sizes $q$ and average degrees $k$. For larger degrees ($k=15$ and $k=50$), the behavior is consistent with the mean-field picture: increasing $\alpha$ reduces the width of the hysteresis region, and the change from a discontinuous to a continuous transition is clearly visible only for $q=3$. However, for $k=5$ and $q=4,5$, the transition is continuous at small $\alpha$, while hysteresis appears only at sufficiently large $\alpha$ and then increases with $\alpha$.}

{To characterize this effect more systematically, Fig.~\ref{fig:hysteresis_PA_2025} shows the hysteresis width as a function of $\alpha$ and $k$ for the model without self-anticonformity and for $q=3$, $q=4$, and $q=5$. The heatmaps demonstrate that the influence of $\alpha$ is non-monotonic and strongly degree-dependent. At small $k$, increasing $\alpha$ can promote hysteresis, whereas for larger $k$ the trend is reversed and hysteresis is suppressed as $\alpha$ increases. Thus, for sparse networks, both $\alpha$ and $k$ determine not only the location of the transition but also its type. This effect is particularly important because it extends the role of $\alpha$ beyond the mean-field borderline case $q=3$: on sparse networks, $\alpha$ can also control the presence or absence of hysteresis for $q=4$ and $q=5$.}

\subsection{\label{Results_empirical_networks} Results on empirical networks}

To test the robustness of our findings beyond idealized random or fully connected graphs, we performed Monte Carlo simulations on empirical social networks. We use the organizational networks collected by Fire \textit{et al.} \cite{fire_organization_2016}. Although we conducted simulations on all networks reported in \cite{fire_organization_2016}, here we present results only for two of them: the small network S2, consisting of $N=320$ nodes with average degree $k=14.81$ and average clustering coefficient $C=0.49$, and the medium network M2 with $N=3862$ nodes, $k=45.22$, and $C=0.31$. Average node degrees of the two selected networks are the closest to Dunbar's number describing sizes of social circles (5, 15, 50 and 150 people who are increasingly distant). Furthermore, structure of these networks lacks hubs (nodes of high degree, whose nodes are sparsely connected with each other), that violate pair approximation assumptions.

Since the fractions of agents with positive expressed and private opinions, $c_S$ and $c_\sigma$, behave qualitatively similarly, in Fig.~\ref{fig:alpha_EPO_networks} we present only the stationary values of $c_S(p)$. We compare Monte Carlo results obtained on empirical networks with simulations on synthetic Watts--Strogatz networks constructed to match the corresponding empirical network in the number of nodes, average degree, and clustering coefficient, as well as with analytical predictions from the pair approximation. 

Results for the small networks (panels a and d in Fig.~\ref{fig:alpha_EPO_networks}) are noisy due to finite-size fluctuations. Despite this, we surprisingly find that for the S2 network the simulations on the empirical graph agree better with the analytical predictions than those on its Watts--Strogatz counterpart. For the M2 network (panels b and e in Fig.~\ref{fig:alpha_EPO_networks}), the results are, first, much smoother. Second, they agree better with the analytical calculations, since this network has a lower clustering coefficient and thus weaker short-range correlations between neighboring nodes. Finally, in the last column of Fig.~\ref{fig:alpha_EPO_networks} we compare the pair-approximation results with simulations on random graphs of size $N=10^4$ and average degree $k=15,50,150$. This network class best matches the pair-approximation assumption of weak correlations between neighbors.

\begin{widetext}

\begin{figure}[htb]
\centering
\includegraphics[width=\linewidth]{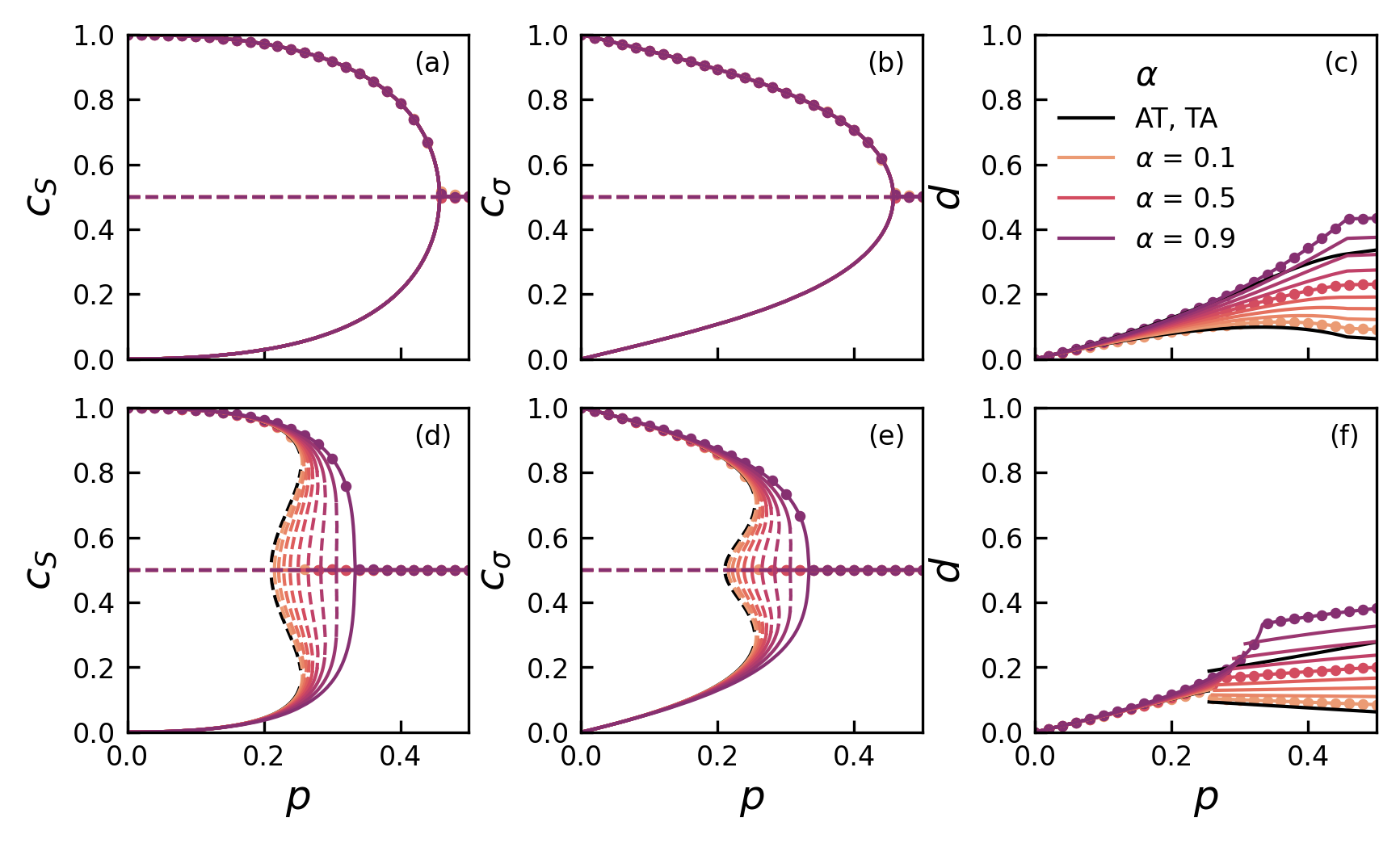}
\caption{\textbf{Stationary concentrations of positive opinions at the expressed level $c_S$ (a, d), private level $c_\sigma$ (b, e), and dissonance $d$ (c, f) as functions of the independence probability $p$, obtained from the mean-field approximation with influence-group size $q=3$.} Panels (a)--(c) correspond to the EPO model with self-anticonformity, whereas panels (d)--(f) correspond to the model without self-anticonformity. Panels (a, d) show results for $c_S$, panels (b, e) for $c_\sigma$, and panels (c, f) for $d$. Results are shown for $\alpha \in \{0.1,0.2,\ldots,0.9\}$, with darker lines corresponding to larger $\alpha$. In panels (a)--(c), the results for all values of $\alpha$ collapse onto the same curves for $c_S$ and $c_{\sigma}$; therefore, only a single line is visible. {Markers represent Monte Carlo simulation results on the complete graph of size $N=10^5$ agents for $\alpha=0.1,0.5,0.9$.}}
\label{fig:alpha_dependence}
\end{figure}

\begin{figure}[htb]
    \centering
    \includegraphics[width=\linewidth]{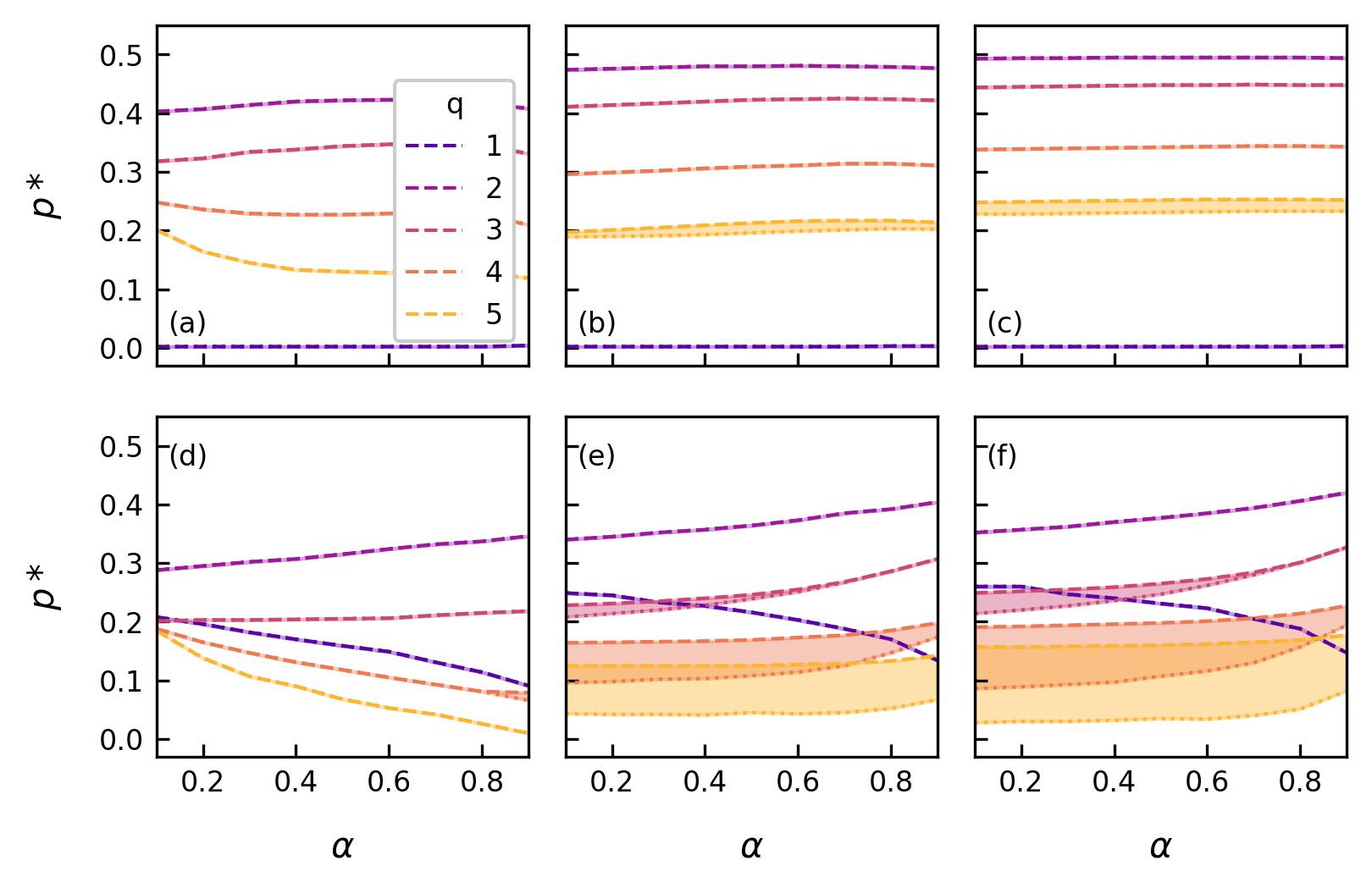}
    \caption{\textbf{Dependence of the lower $p^*_{low}$ (dotted lines) and upper $p^*_{up}$ (dashed lines) spinodals on $\alpha$, obtained from the pair approximation.} Panels (a)--(c) correspond to the EPO model with self-anticonformity, whereas panels (d)--(f) correspond to the model without self-anticonformity. Panels (a, d) show results for networks with average degree $k=5$, panels (b, e) for $k=15$, and panels (c, f) for $k=50$. {Different sizes of the influence group $q$ are indicated by the common legend in panel (a). The shaded region between the lines indicates hysteresis.}}
    \label{fig:spinodals_PA}
\end{figure}

\begin{figure}[htb]
    \centering
    \includegraphics[width=\linewidth]{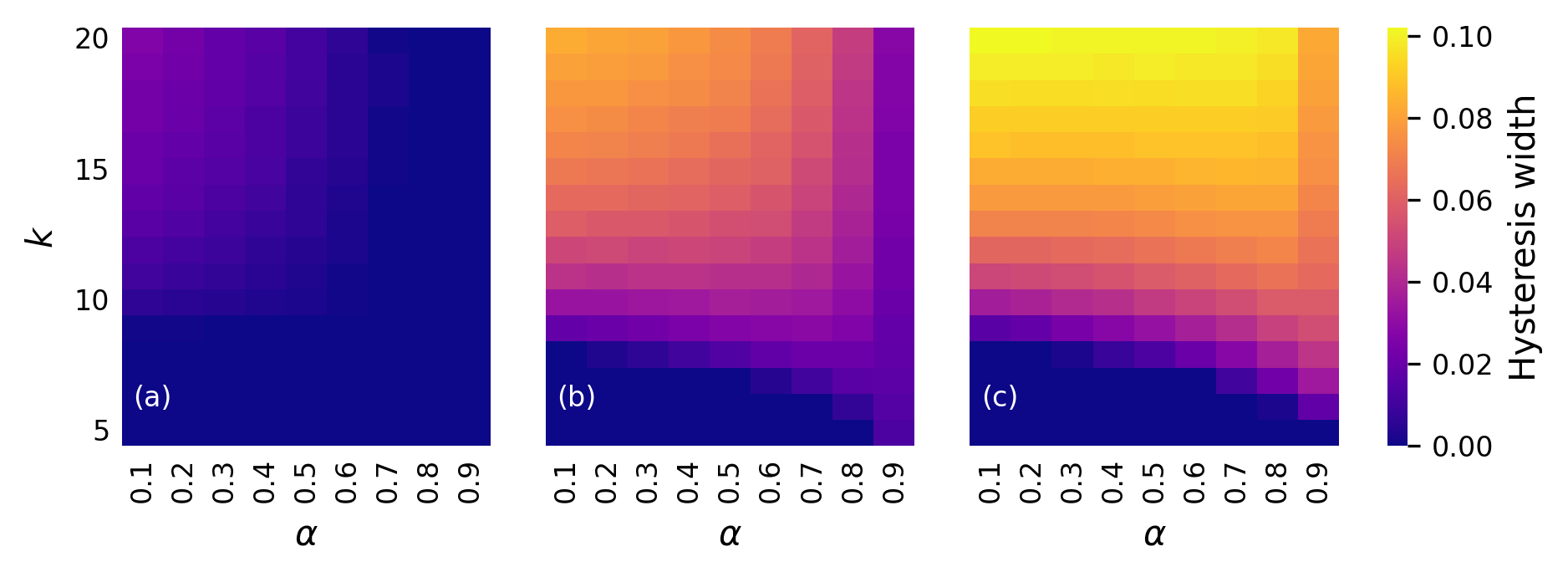}
    \caption{{\textbf{Dependence of the hysteresis width on $\alpha$ and $k$.} Results are shown for the model without self-anticonformity. The hysteresis width, obtained within the pair approximation as the difference between the upper and lower spinodals, is shown for (a) $q=3$, (b) $q=4$, and (c) $q=5$. The lighter color indicates a wider hysteresis region, as shown by the colorbar.}}
    \label{fig:hysteresis_PA_2025}
\end{figure}

\begin{figure}[htb]
    \centering
    \includegraphics[width=\linewidth]{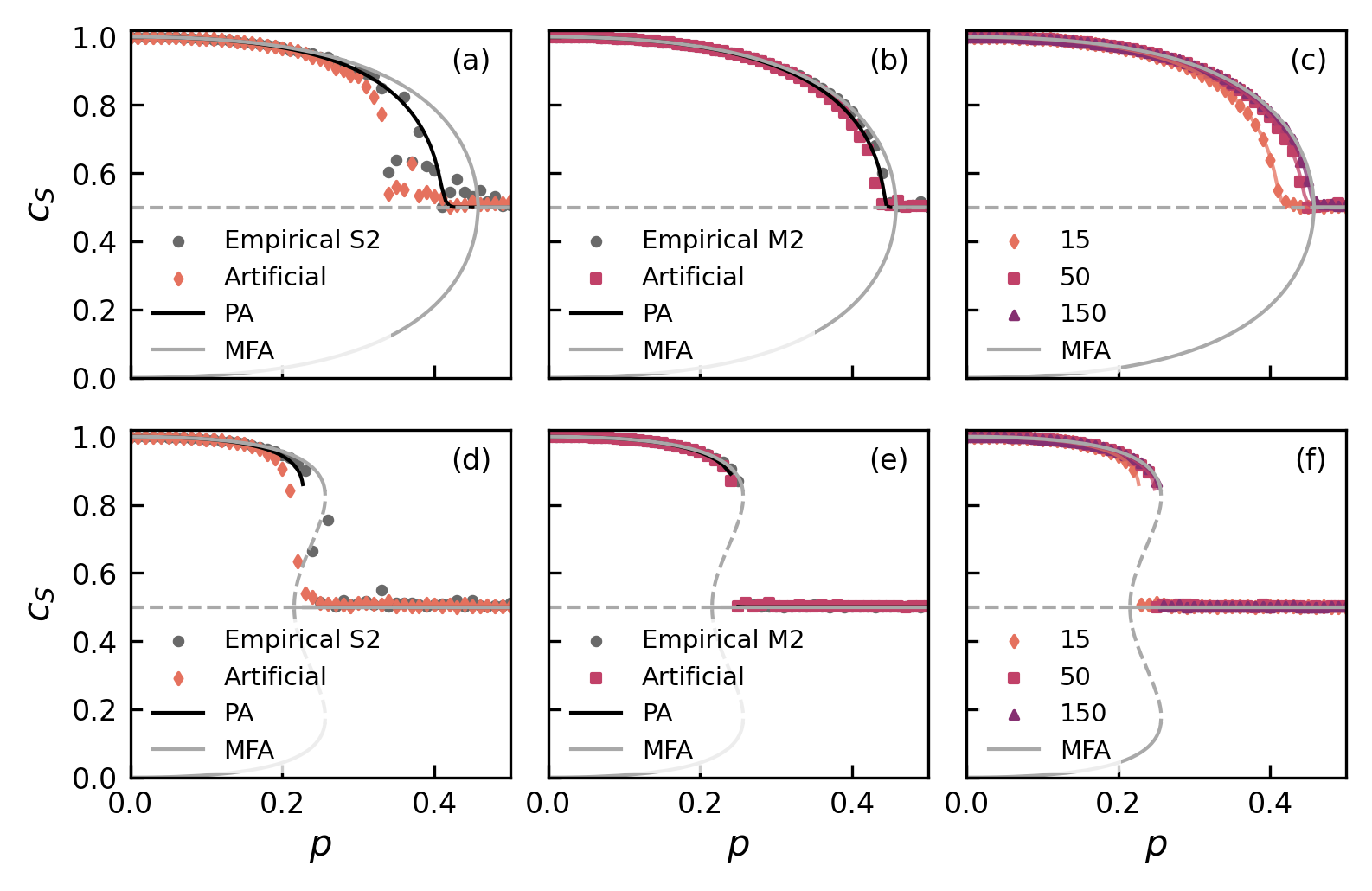}
    \caption{\textbf{Stationary concentration of positive expressed opinions $c_S$ as a function of the independence probability $p$ for $\alpha=0.1$ and $q=3$.} Monte Carlo results on empirical networks~\cite{fire_organization_2016} are compared with simulations on Watts--Strogatz networks matched in size, average degree, and clustering coefficient in panels (a), (b), (d), and (e). Panels (c) and (f) show simulations on random graphs of size $N=10^4$ with average degrees $k=15$, $k=50$, and $k=150$. Panels (a)--(c) correspond to the EPO model with self-anticonformity, whereas panels (d)--(f) correspond to the model without self-anticonformity. Full markers correspond to ordered initial conditions, while empty markers correspond to random initial conditions. Simulation results are compared with the pair-approximation predictions.}
    \label{fig:alpha_EPO_networks}
\end{figure}

\end{widetext}

\section{\label{sec:Conclusions}Conclusions}
In this paper, we introduced and compared two versions of the $\alpha$-EPO model, extending the two $q$-voter EPO variants proposed in~\cite{jedrzejewski_think_2018, kaminska_impact_2025}. These two variants differ in that one includes self-anticonformity whereas the other does not. A parallel analysis of the corresponding $\alpha$-EPO versions allows us to investigate the role of self-anticonformity, which has recently attracted significant attention~\cite{dvorak_strategic_2024, lipiecki_depolarizing_2025}. We modified the original models by allowing internal and public states to evolve at different rates, which is motivated by the fact that people's private and expressed opinions, beliefs and attitudes, or intentions and behaviors do not necessarily change simultaneously or equally often.

Both $\alpha$-EPO variants display a transition between an ordered (agreement) phase and a disordered (disagreement) phase. Although the position of the critical point $p^*$ and the character of the transition depend on the variant, within each model the expressed and private order parameters, $c_S(p)$ and $c_\sigma(p)$, undergo the transition at the same $p^*$ and with the same transition type. Importantly, in the ordered phase ($p<p^*$) agreement is systematically stronger at the expressed than at the private level, indicating pluralistic ignorance: public consensus can overstate the true degree of private agreement because agents misperceive the prevailing norm and adjust their expressed opinions accordingly.

In the variant with self-anticonformity, we find that the model is robust with respect to $\alpha$ as long as $0<\alpha<1$. Specifically, the stationary fractions of agents with positive expressed and private opinions, $c_S(p)$ and $c_\sigma(p)$, are identical for all $\alpha$. The only quantity that increases with $\alpha$ is the fraction of agents in dissonance, $d$, i.e., agents whose expressed and private opinions disagree. Hence, from an individual perspective, it may be beneficial for private opinions to be less volatile since fewer agents experience cognitive dissonance. At the same time, from a collective perspective, the update frequency does not matter because the stationary macroscopic state remains unchanged.

Once self-anticonformity is excluded, this robustness disappears: the stationary dependencies $c_S(p)$ and $c_\sigma(p)$ change with $\alpha$. In this variant, our results approach those of the original model~\cite{kaminska_impact_2025} in the limit $\alpha \to 0$, which can be understood intuitively from the update rules: when updates occur predominantly at the expressed level, the expressed layer relaxes faster, and private conformity events are more likely to be effective because they are permitted only when the target's expressed opinion already matches the unanimous influence group. Within the mean-field approximation, for influence-group sizes $q \geq 2$, the critical point $p^*$ increases with $\alpha$, meaning that for larger $\alpha$ it is easier to reach an agreement; a majority of agents choose the same option even for a higher probability of independence $p$.

{The effect of $\alpha$ on the type of phase transition depends, however, on the interaction topology. In the mean-field approximation, corresponding to the complete-graph limit, increasing $\alpha$ narrows the hysteresis region and, for $q=3$, leads to its disappearance at a critical value $\alpha^*$, indicating a change from a discontinuous to a continuous phase transition. Thus, in the complete-graph limit, more frequent private-opinion updates weaken bistability and reduce path dependence. The pair approximation shows that this conclusion must be refined for sparse networks. At small average degree $k$, the hysteresis width can increase with $\alpha$, which is opposite to the trend observed for larger $k$ and in the mean-field limit. Moreover, in this low-connectivity regime, $\alpha$ and $k$ jointly determine the type of phase transition also for larger influence groups. Therefore, the emergence or suppression of hysteresis is controlled not by $\alpha$ alone, but by the combined effect of update-rate asymmetry, influence-group size, and network connectivity.}

We are aware that the above conclusions should be treated with caution, as they are derived from a specific class of models. {We therefore indicate two important directions for future work. First, the same idea of unequal update rates could be applied to other models with private and expressed opinions. This would help assess how robust the role of $\alpha$ is across different formulations of EPO dynamics and network structures. Second, incorporating external sources of influence, such as mass media, political parties, campaigns, or advertisements, into the proposed model is a natural and important next step toward a more holistic computational framework~\cite{doniec_2025}. We deliberately leave this extension for future work, because the aim of the present study is to isolate the effect of unequal update rates between private and expressed opinions, controlled by the parameter $\alpha$. Adding external fields would considerably enlarge the parameter space and could obscure this central mechanism. Moreover, external influence can be introduced into an EPO model in several non-equivalent ways: it may act on the expressed layer, on the private layer, on both layers simultaneously, or with different strengths and even different directions at the two levels. Since these alternatives may generate qualitatively different dynamics, they require a separate systematic analysis. Thus, the present work should be viewed as a baseline study that identifies the dynamical role of unequal update rates, while providing a clear foundation for future extensions that include structured external influence.}
\begin{acknowledgments}
Funded by the National Science Centre, Poland under the OPUS call in the Weave programme, project no. 2023/51/I/HS6/02269.
\end{acknowledgments}

\section*{Data Availability Statement}
The source codes in C++ and MATLAB, which allow for the reproduction of all results in this publication, are publicly available on \href{https://github.com/BarbaraKaminska/EPO-q-voter}{https://github.com/BarbaraKaminska/EPO-q-voter}. The empirical network data used in this study is sourced from \url{https://data4goodlab.github.io/dataset.html}.

\appendix
\section{Stationary concentrations for $q=2$ and $q=4$}
\label{sec:appendix}
\begin{widetext}

\begin{figure*}[h]
\centering
\includegraphics[width=\linewidth]{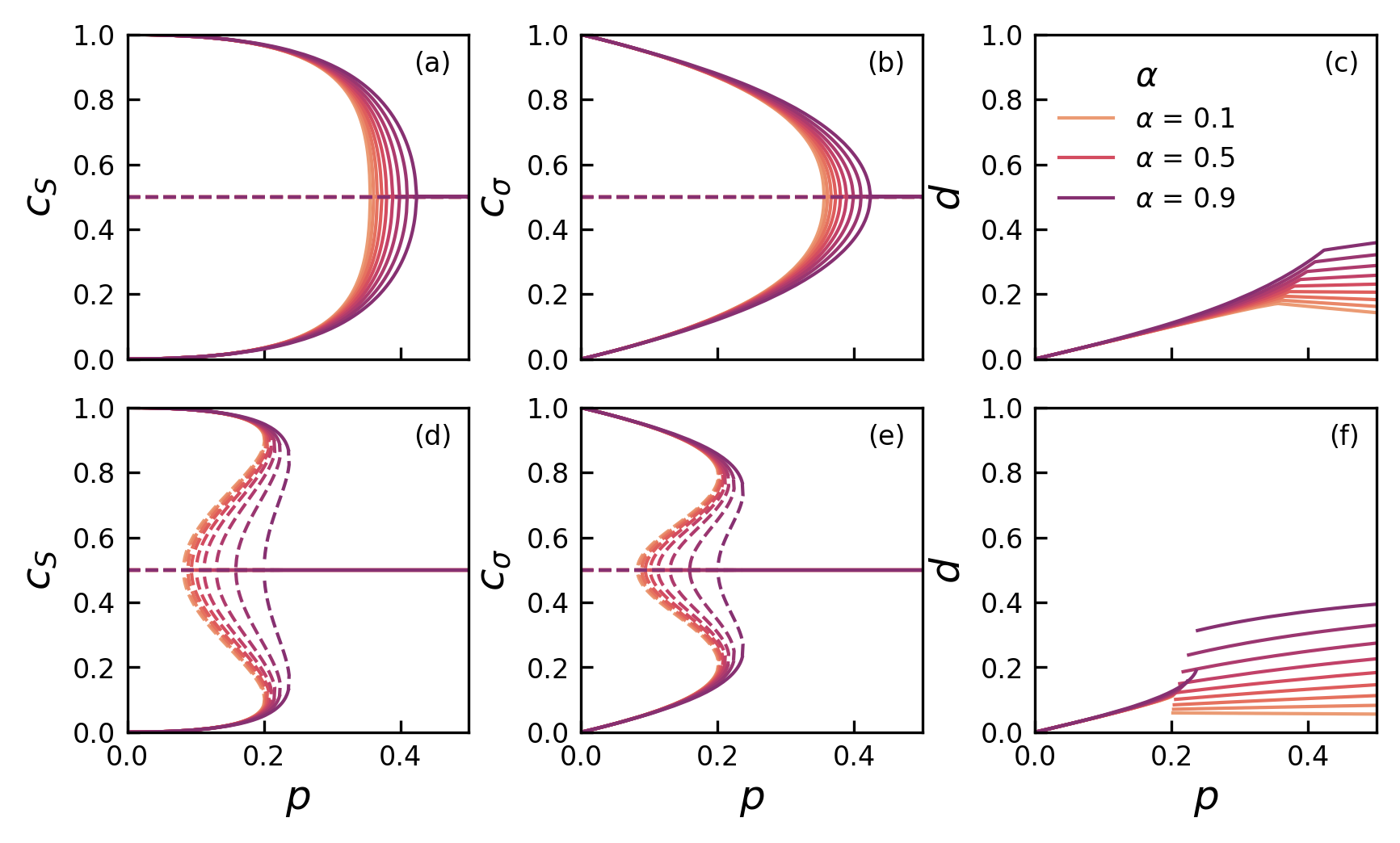}
\caption{{\textbf{Stationary concentrations of positive opinions at the expressed level $c_S$ (a, d), private level $c_\sigma$ (b, e), and dissonance $d$ (c, f) as functions of the independence probability $p$, obtained from the mean-field approximation for the model without self-anticonformity.} Panels (a)--(c) correspond to influence-group size $q=2$, whereas panels (d)--(f) correspond to $q=4$. Panels (a, d) show results for $c_S$, panels (b, e) for $c_\sigma$, and panels (c, f) for $d$. Results are shown for $\alpha \in \{0.1,0.2,\ldots,0.9\}$, with darker lines corresponding to larger $\alpha$. For clarity, the legend displays only selected values of $\alpha$.}}
\label{fig:alpha_dependence_q_2_4}
\end{figure*}

\end{widetext}

{
To complement the discussion of the borderline case $q=3$, we present in Fig.~\ref{fig:alpha_dependence_q_2_4} representative mean-field results for the model without self-anticonformity for $q=2$ and $q=4$.
These cases illustrate that the qualitative change in the order of the transition induced by increasing $\alpha$ is specific to $q=3$. For $q=2$, the transition remains continuous for all considered values of $\alpha$, whereas for $q=4$ it remains discontinuous. Changing $\alpha$ shifts the stationary curves and the transition point, but it does not change the type of transition in these two cases.}

\FloatBarrier

\bibliography{references}

\end{document}